\documentclass[final,3p,times]{elsarticle}

\usepackage{amssymb}
\usepackage{amsmath}

\journal{Journal of Subatomic Particles and Cosmology}

\begin{document}

\begin{frontmatter}

\title{Momentum anisotropy from Resistive Magnetohydrodynamics}

\author[aaa]{Khwahish Kushwah}
\ead{khwahish.kushwah@id.uff.br}

\author[aaa]{Gabriel Denicol}
\ead{gsdenicol@id.uff.br}

\affiliation[aaa]{organization={Instituto de F\'isica, Universidade Federal Fluminense},
             addressline={Praia Vermelha campus},
             city={Niter\'oi},
             postcode={24210-346},
             state={RJ},
             country={Brazil}}

\begin{abstract}
We derive relativistic resistive magnetohydrodynamics framework for a two-component ultrarelativistic plasma of massless, oppositely charged particles directly from the Boltzmann--Vlasov equation using the 14-moment approximation. The resulting second-order equations couple the net-charge diffusion current to the shear-stress tensor through the electric field, with all transport coefficients given in closed microscopic form. In the homogeneous limit, the charge-current dynamics is well described by relaxation-type Ohm's law for moderate field strengths, while large viscosity drives the system into an underdamped oscillatory regime absent from the standard Israel--Stewart formulation. Most strikingly, a purely electric field generates sizable momentum anisotropy even without any underlying flow gradient. Under Bjorken expansion this field-induced anisotropy persists but becomes subleading to the hydrodynamic expansion source.
\end{abstract}

\begin{keyword}
Momentum anisotropy, resistive magnetohydrodynamics, dissipative hydrodynamics, kinetic theory.
\end{keyword}

\end{frontmatter}

\paragraph{\textbf{Introduction}}  Relativistic plasmas in strong electromagnetic fields arise in compact astrophysical systems, the early Universe, and high-energy heavy-ion collisions~\cite{Skokov:2009qp}. In such systems, electromagnetic fields can significantly affect the evolution of the medium, requiring a consistent coupling between hydrodynamics and Maxwell's equations~\cite{Bjorken1983}. While relativistic magnetohydrodynamic frameworks have been developed also for strong fields, a kinetic-theory derivation of resistive MHD for a two-component plasma remains important for identifying the microscopic origin of charge transport and its coupling to viscous dynamics.

Here we summarize such a derivation for a plasma of massless, oppositely charged particles~\cite{Kushwah:2025rmhd}, extending our earlier magnetic-field-only treatment~\cite{Kushwah:2024mhd}. Starting from the Boltzmann--Vlasov equation and using the 14-moment approximation in the Landau frame, we obtain coupled evolution equations for the charge-diffusion current and the shear-stress tensor. In homogeneous systems, the electric field alone sources momentum anisotropy, even in the absence of flow gradients, while nonlinear corrections to Ohm's law become relevant only for sufficiently strong fields. We present the closed equations, microscopic transport coefficients, and numerical solutions for homogeneous matter and Bjorken flow.

\section{Kinetic framework and the closed equations of motion}
\label{sec:framework}

\noindent We consider a relativistic dilute gas of massless, oppositely charged particles,
with single-particle distribution functions $f^{\pm}_k$, where $k^\mu$ is the
four-momentum and the $\pm$ labels the charge species. Each species obeys the
Boltzmann--Vlasov equation,
\begin{equation}
    k^{\mu}\partial_{\mu}f^{\pm}_k \pm |q|\,k_{\nu}F^{\mu\nu}
    \frac{\partial f^{\pm}_k}{\partial k^{\mu}}
    = C[f^{\pm}_k,f^{\pm}_{k'}] + C[f^{\pm}_k,f^{\mp}_{k'}],
    \label{eq:BV}
\end{equation}
where $F^{\mu\nu}$ is the electromagnetic field tensor, $|q|$ is the particle charge,
and the right-hand side is the collision kernel accounting for both intra-species and inter-species 
binary elastic scatterings, characterized by constant transport cross sections
$\sigma_T^{++}=\sigma_T^{--}\equiv\sigma_T$ and $\sigma_T^{+-}=\sigma_T^{-+}$.

The macroscopic fields are defined by decomposing the energy-momentum tensor
$T^{\mu\nu}_\pm$ and particle four-current $N^\mu_\pm$ of each species with respect
to the fluid four-velocity $u^\mu$ (normalized as $u_\mu u^\mu = 1$). Landau
matching fixes the frame via $u_{\mu}\!\left(T^{\mu\nu}_++T^{\mu\nu}_-\right) \equiv \epsilon_0\, u^{\nu}$, and $
    u_{\mu}\!\left(N^{\mu}_+-N^{\mu}_-\right) \equiv n_0$, where $\epsilon_0$ is the total energy density and $n_0$ the net charge number
density in the local rest frame. From these one defines the net-charge diffusion
four-current $V^{\mu}_q = |q|(V^{\mu}_+-V^{\mu}_-)$, where $V^\mu_\pm$ are the
individual species diffusion currents, and the total shear-stress tensor
$\pi^{\mu\nu}=\pi^{\mu\nu}_++\pi^{\mu\nu}_-$, which characterizes momentum
anisotropy in the local rest frame.

Closure of the moment hierarchy is achieved through the 14-moment ansatz
\cite{Israel:1979wp},
$f^{\pm}_k = f^{\pm}_{0k}\!\left(1+\varphi^{\pm}_k\right)$, such that $ \varphi^{\pm}_k \approx -\frac{5\,k_{\mu}V^{\mu}_{\pm}}{n_0 T} +\frac{k_{\mu}k_{\nu}\pi^{\mu\nu}_{\pm}}{2(\epsilon_0+P_0)T^2}$, where $f^{\pm}_{0k}$ is the local-equilibrium distribution, $T$ is the temperature,
and $P_0$ is the equilibrium pressure. Taking moments of Eq.~\eqref{eq:BV},
combining the two species, and eliminating the 
relative energy-diffusion current $\delta h^{\mu}\equiv h^{\mu}_+-h^{\mu}_-$,
with $h^\mu_\pm$ being the individual heat flows, through a hydrodynamic
gradient expansion \cite{Kushwah:2025rmhd}, yields a closed, coupled set of
second-order equations of motion. For a locally neutral plasma ($n_q=0$) in the
absence of a magnetic field ($B=0$), they read
\begin{align}
\tau_{V_q}\dot{V}_q^{\langle\mu\rangle} + \Gamma_{V_q}V^{\mu}_q
    &= G_E E^{\mu}
      + \Omega_{E\pi}E_{\nu}\pi^{\mu\nu}
      - \tau_{V_q}\theta\, V^{\mu}_q
      + c_{\sigma}\sigma^{\mu}_{\ \nu}V^{\nu}_q
      - \Gamma_{\rm NL}(E\!\cdot\! V_q)V^{\mu}_q,
\label{eq:Vq-full}\\
\tau_{\pi}\dot{\pi}^{\langle\mu\nu\rangle} + \pi^{\mu\nu}
    &= \frac{8}{5}\tau_{\pi}V_q^{\langle\mu}E^{\nu\rangle}
      + \frac{8}{15}\epsilon\,\sigma^{\mu\nu}
      - \frac{4}{3}\theta\,\pi^{\mu\nu}
      - \frac{10}{7}\sigma^{\lambda\langle\mu}\pi^{\nu\rangle}_{\lambda},
\label{eq:pi-full}
\end{align}
where $E^\mu$ is the electric four-field in the fluid rest frame, $\theta\equiv\nabla_\mu u^\mu$
is the expansion scalar, $\sigma^{\mu\nu}\equiv\nabla^{\langle\mu}u^{\nu\rangle}$
is the shear tensor, and an overdot denotes the comoving derivative
$\dot{A}^\mu\equiv u^\nu\nabla_\nu A^\mu$. Angle brackets on indices denote the
symmetric, traceless projection orthogonal to $u^\mu$. And according to the Einstein summation convention, $A \cdot B = A^\mu B_\mu$.

The source term $V_q^{\langle\mu}E^{\nu\rangle}$ in Eq.~\eqref{eq:pi-full} is
the key structural result of the derivation. It shows that an electric field
couples \emph{directly} to the shear-stress tensor, sourcing momentum anisotropy
independently of any velocity gradient. Conversely, $\pi^{\mu\nu}$ feeds back on
the charge-current relaxation through $\Omega_{E\pi}\,E_{\nu}\,\pi^{\mu\nu}$ in
Eq.~\eqref{eq:Vq-full}, closing a nonlinear loop between electromagnetic and
viscous degrees of freedom.

Every transport coefficient is a closed-form function of the microscopic cross
sections and the particle number density $n$. The transport coefficients appearing in Eqs.~\eqref{eq:Vq-full}--\eqref{eq:pi-full} 
are all given in closed microscopic form: $\tau_{V_q} = (1-\alpha_{V_q})/\Gamma_{V_q}$, 
$\alpha_{V_q} = \sigma_T/(9\sigma_T^{+-})$, 
$\Gamma_{V_q} = n(2\sigma_T^{+-}+5\sigma_T/6)/3$, 
$G_E = |q|^2 n(1+\sigma_T/3\sigma_T^{+-})/(3T)$, 
$\sigma_E = G_E/\Gamma_{V_q}$, 
$\Omega_{E\pi} = |q|^2/(20T^2\Gamma_{V_q})$, 
$\Gamma_{\rm NL} = \sigma_T/(27T\sigma_T^{+-}n\Gamma_{V_q})$, 
and $\Sigma_\pi \equiv 1/\tau_\pi = 3n(\sigma_T^{+-}+\sigma_T)/10 = (\epsilon+P_0)/5\eta$, 
the last equality identifying the shear viscosity $\eta$ in the field-free limit and
$c_\sigma$ follows analogously from the gradient expansion~\cite{Kushwah:2025rmhd}.

\section{Homogeneous case}
\label{sec:homogeneous}
\noindent To isolate the role of the electric field, we consider a spatially homogeneous
system, $\theta=\sigma^{\mu\nu}=0$, which is locally neutral and satisfies
$E^\mu \parallel x^\mu$ and $B=0$. In this limit, all velocity-gradient terms in Eqs.~\eqref{eq:pi-full}--\eqref{eq:Vq-full}
vanish identically and all dissipative currents are taken to be initially zero,
so any departure from equilibrium is driven solely by $E^\mu$. The system then
closes through Maxwell's equation, $\dot{E}^{\mu}=-V_q^\mu $ which depletes the field as the charge current builds up. This controlled setting therefore provides a clean test of
whether an electric field alone, with no underlying flow profile, can generate
both a net-charge current and, through the
$V_q^{\langle\mu}E^{\nu\rangle}$ source in Eq.~\eqref{eq:pi-full}, a nonzero
shear-stress tensor.

\begin{figure}[!htb]
\centering
\begin{minipage}{0.48\columnwidth}
    \centering
    \includegraphics[width=\linewidth]{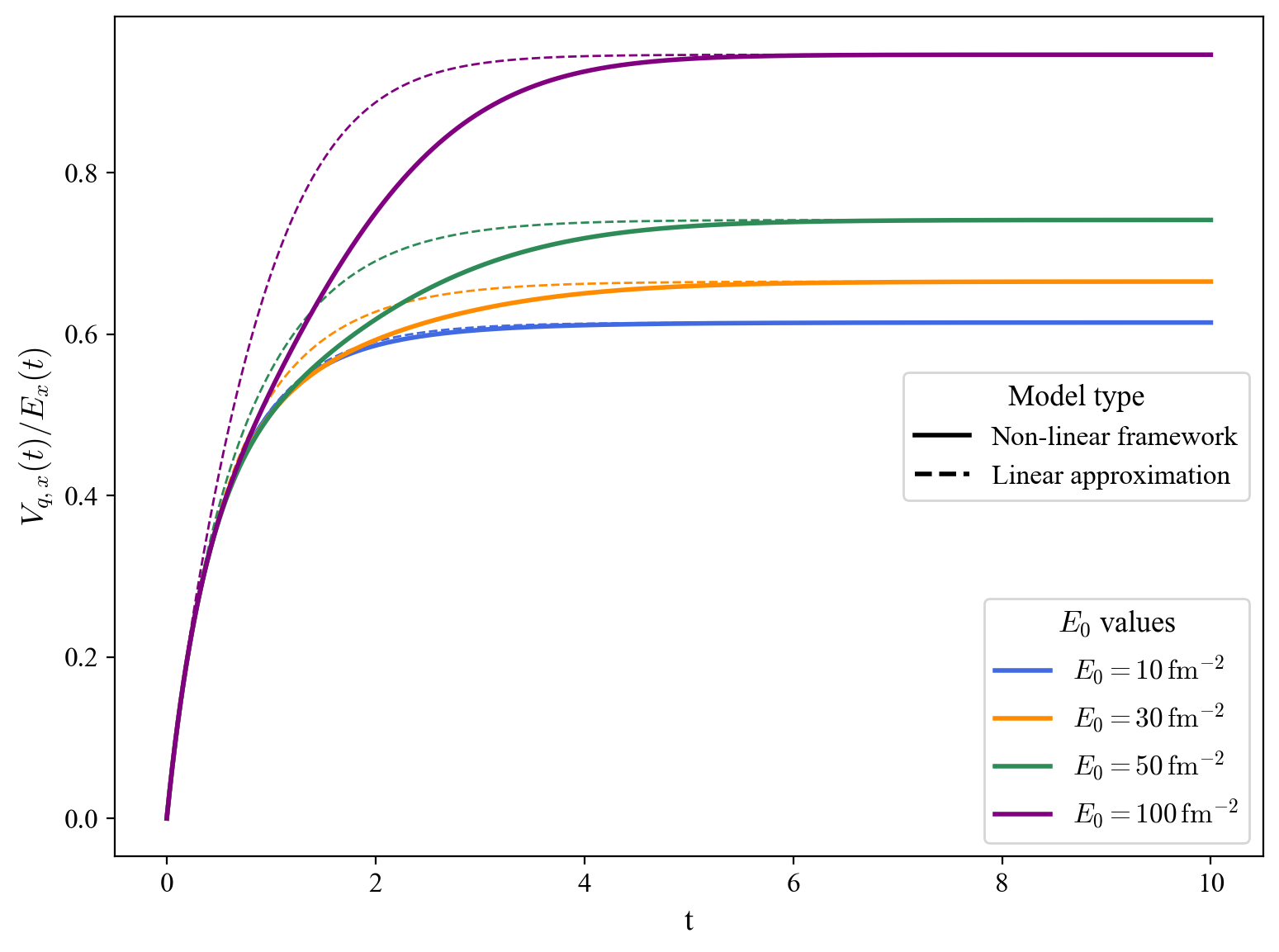}
    \caption{Normalized charge current $V_{q,x}(t)/E_x(t)$ for $\eta/s=1$,
    $\sigma_T/\sigma_T^{+-}=0.1$, and several $E_0$, comparing the full
    nonlinear evolution (solid) against the linearized Ohm's law (dashed).
    Both converge to the same Ohmic asymptote at late times.}
    \label{fig:2}
\end{minipage}
\hfill
\begin{minipage}{0.48\columnwidth}
    \centering
    \includegraphics[width=\linewidth]{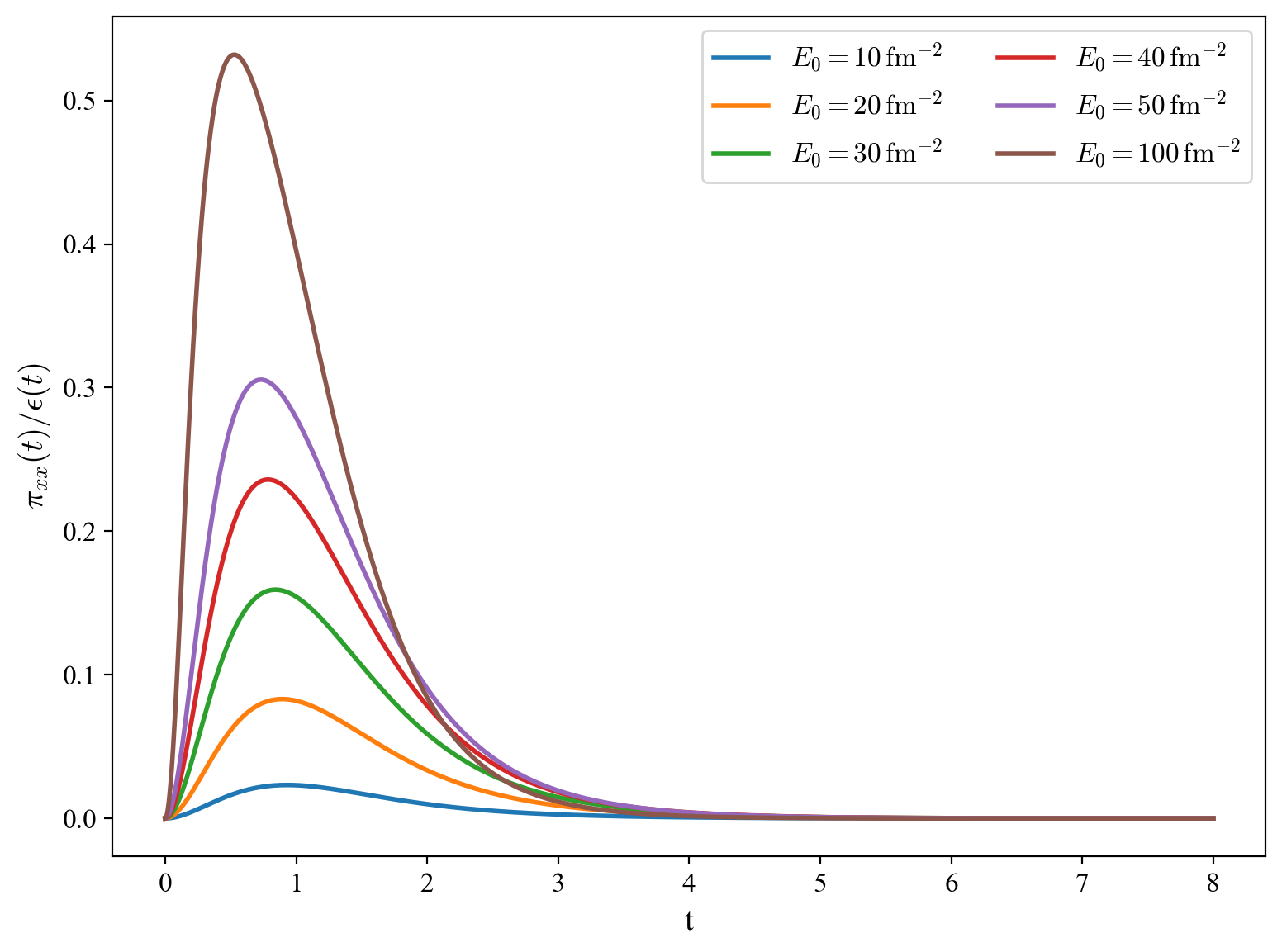}
    \caption{Momentum anisotropy $\pi_{xx}/\epsilon$ generated solely by the
    electric field, for $\eta/s=1$, $\sigma_T/\sigma_T^{+-}=0.1$, and several
    $E_0$. The anisotropy grows with $E_0$ and vanishes once the field is
    depleted by the charge current.}
    \label{fig:4}
\end{minipage}
\end{figure}

\paragraph{\textit{\textbf{a. Ohm's law and its nonlinear corrections}}} Further switching off $\Omega_{E\pi}$ and $\Gamma_{\rm NL}$ in Eq.~\eqref{eq:Vq-full} reduces the charge-current equation to the familiar relaxation-type Ohm's law, $\tau_{V_q}\dot V^{\mu}_q+V^{\mu}_q=\sigma_E E^{\mu}$, which we use as the baseline (``linear approximation'') against which the full nonlinear theory is compared. Figure~\ref{fig:2} shows the resulting normalized current $V_{q,x}(t)/E_x(t)$ for $\eta/s=1$, $\sigma_T/\sigma_T^{+-}=0.1$, and several $E_0$: at late times, both the linear and nonlinear curves converge to a common, field-independent asymptote, confirming that the steady-state conductivity remains Ohmic and that the late-time decay of the current simply tracks the decaying electric field. Overall, the dynamics of the net-charge current is well described by the conventional linearized Ohm's law for $E_0\lesssim30$~fm$^{-2}$, with nonlinear corrections only becoming relevant beyond this threshold.

\paragraph{\textbf{\textit{b. Momentum anisotropy}}} Figure~\ref{fig:4} shows the numerical solution of $\pi_{xx}(t)/\epsilon(t)$ for $\eta/s=1$, $\sigma_T/\sigma_T^{+-}=0.1$ and several $E_0$: the electric field, acting solely through the $V_q^{\langle\mu}E^{\nu\rangle}$ source of shear stress tensor, produces a transient but sizable momentum anisotropy that grows with $E_0$ and relaxes back to zero once the field is depleted by the charge current, restoring isotropy at late times. This is the central result of our analysis: a purely electric field, without any underlying velocity gradient, can anisotropize the local momentum distribution of the plasma. The effect is largest at early times, of order $E_0\sim20$~fm$^{-2}$ (corresponding to $\sim10^{19}$~Gauss) being sufficient to generate $\pi_{xx}/\epsilon\sim\mathcal{O}(0.1)$, comparable to typical shear corrections from flow gradients alone in heavy-ion collisions.

\section{Numerical solution and Bjorken flow}
\label{sec:numerics}

Equations~\eqref{eq:Vq-full}--\eqref{eq:pi-full} along with the Maxwell's equations were solved numerically in \texttt{Python}, as a first-order ODE system integrated with \texttt{scipy.integrate.solve\_ivp} (adaptive Runge-Kutta, RK45), starting from an equilibrium state of energy density $\epsilon_0=1000$~fm$^{-4}$, for $\eta/s\in\{0.5,1,5\}$ and $E_0\in[10,100]$~fm$^{-2}$.

To test whether the momentum anisotropy of Sec.~\ref{sec:homogeneous} survives in a more realistic, rapidly expanding setting, we further solve Eqs.~\eqref{eq:Vq-full}--\eqref{eq:pi-full} for the boost-invariant Bjorken profile, $\tau=\sqrt{t^2-z^2}$, $\xi=\frac{1}{2}\ln[(t+z)/(t-z)]$, $g_{\mu\nu}={\rm diag}(1,-1,-1,-\tau^2)$, for which $\theta=1/\tau$ and $\sigma^{\mu\nu}={\rm diag}(0,\frac{1}{3\tau},\frac{1}{3\tau},-\frac{2}{3\tau^2})$ in Milne coordinates. This turns Eqs.~\eqref{eq:Vq-full}--\eqref{eq:pi-full} into an explicit ODE system in $\tau$ for $V_q^x$, $\pi^{xx}$, $\pi^{\eta\eta}$, $E^x$, e.g.
\begin{equation}
\dot\pi^{xx} + \Sigma_{\pi}\pi^{xx} = \frac{16}{15}V_{q,x}E_x + \frac{8}{45}\frac{\epsilon}{\tau} - \frac{14}{9}\frac{\pi^{xx}}{\tau}, \qquad
\dot E^x + \frac{E^x}{\tau} = -V_q^x,
\end{equation}
closed with 
\begin{equation}
    \dot\epsilon=-\frac{4}{3\tau}\epsilon+\frac{1}{3\tau}(\pi^{xx}+\pi^{yy}-2\tau^2\pi^{\eta\eta})+E^xV_q^x.
\end{equation}
We integrate from $\tau_0=0.1$~fm with $\epsilon_0=1000$~fm$^{-4}$, $E_0\in[10,100]$~fm$^{-2}$, again comparing the nonlinear and linearized (Ohm's law) systems.

\begin{figure}[!htb]
\centering
\includegraphics[width=\columnwidth]{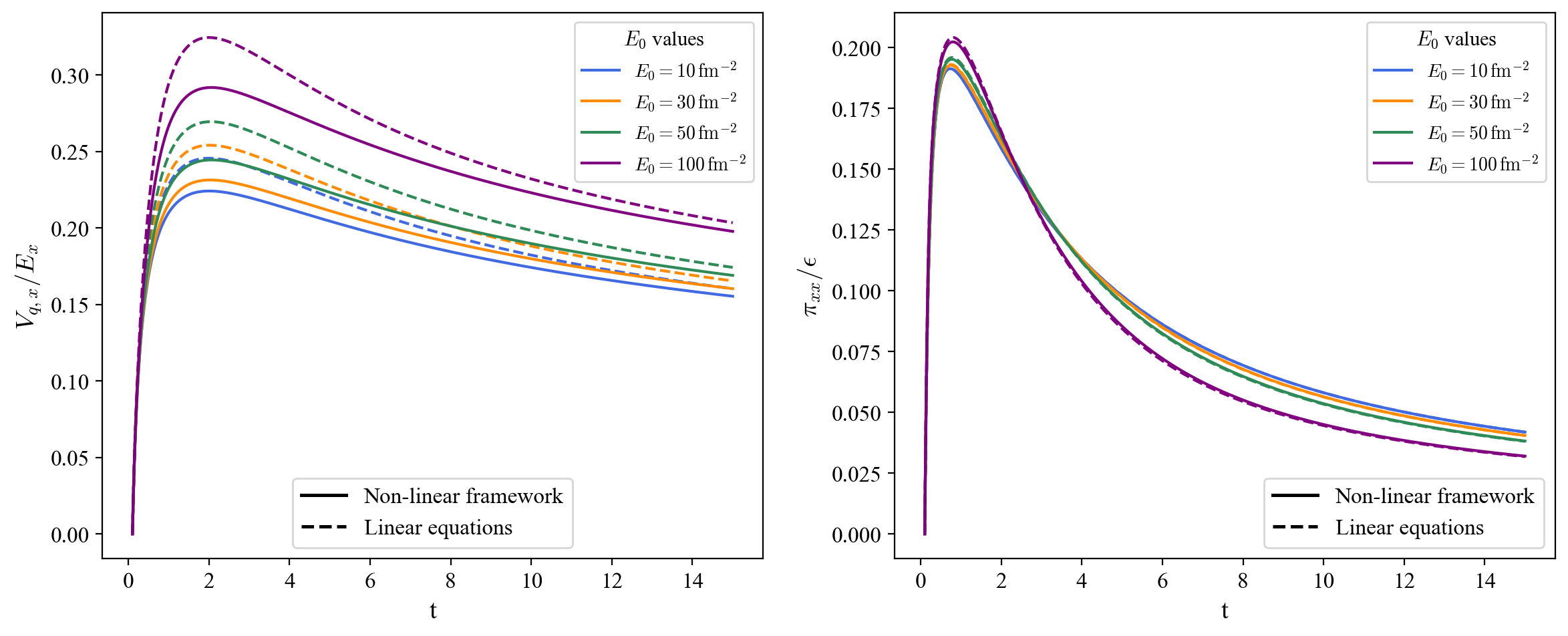}
\caption{Bjorken flow: numerical solution of $V_{q,x}/E_x$ (left) and $\pi_{xx}/\epsilon$ (right) versus $\tau$, for several $E_0$, comparing the nonlinear (solid) and linearized Ohm's law (dashed) systems.}
\label{fig:5}
\end{figure}

As shown in Fig.~\ref{fig:5}, $V_{q,x}$ is still generated by the electric field but decays faster than in the homogeneous case, since $E^x$ is itself diluted by the $1/\tau$ term in Maxwell's equation. The linearized Ohm's law remains a reasonable approximation to this decay for moderate field strengths, but departs visibly from the full nonlinear result once $E_0$ becomes large, mirroring the homogeneous-case breakdown of Sec.~\ref{sec:homogeneous} but now compounded by the expansion rate. The momentum anisotropy, by contrast, becomes dominated by the $8\epsilon/(45\tau)$ expansion source rather than by $E_0$: $\pi_{xx}/\epsilon$ shows an early peak essentially independent of $E_0$, with the electric field only mildly accelerating its late-time decay. The diffusion-shear coupling identified in Sec.~\ref{sec:homogeneous} thus persists under longitudinal expansion, but its contribution to the momentum anisotropy becomes subleading to the hydrodynamic expansion itself.

\section{Conclusions}
\label{sec:conclusions}

\noindent We derived, from the Boltzmann--Vlasov equation for a two-component massless plasma, closed second-order equations coupling the charge-diffusion current to the shear-stress tensor via the electric field, with all transport coefficients given in closed microscopic form. The key result is the source term $V_q^{\langle\mu}E^{\nu\rangle}$ in the shear equation, showing that an electric field alone can generate momentum anisotropy without any flow gradient.

In the homogeneous case, the charge current follows the linearized Ohm's law up to $E_0\lesssim30$~fm$^{-2}$ (Fig.~\ref{fig:2}), while large viscosity can push the system into an oscillatory regime beyond standard Israel--Stewart. The electric field alone produces a sizable, transient anisotropy $\pi_{xx}/\epsilon\sim\mathcal{O}(0.1)$ for $E_0\sim20$~fm$^{-2}$ (Fig.~\ref{fig:4}), comparable to typical flow-driven shear effects. Under Bjorken expansion (Fig.~\ref{fig:5}), this coupling persists, but the anisotropy becomes dominated by the expansion source rather than $E_0$, and the oscillatory regime is not reached.

These results identify, from kinetic theory, a direct microscopic channel through which electromagnetic fields anisotropize a relativistic plasma independently of flow gradients, thus motivating future extensions to finite and large magnetic field and net charge density.

\bibliographystyle{elsarticle-num}
\bibliography{sqm2026_template}

\end{document}